\documentclass[aps,prb,twocolumn,groupedaddress]{revtex4-1} 

\usepackage{graphicx}
\usepackage{dcolumn}
\begin{document}

\title{Phenomenological theory of optical broadening in zero-dimensional systems applied to silicon nanocrystals}
\author{V.~V.~Nikolaev}
\email[]{valia.nikolaev@gmail.com}
\author{N.~S.~Averkiev}
\affiliation{Ioffe Institute, St. Petersburg 194021, Russia}
\author{Minoru Fujii}
\affiliation{Department of Electrical and Electronic Engineering,
Graduate School of Engineering, Kobe University,
Kobe 657-8501, Japan}

\date{\today}

\begin{abstract}
We develop a phenomenological theory of inhomogeneous broadening in
zero-dimensional systems and apply it to study
photoluminescence (PL) spectra of silicon nanocrystals measured at helium and
room temperatures. The
proposed approach allowed us to explain experimentally observed PL
peak asymmetry, linear dependence of the peak width on its maximum
and anomalous alteration of spectral characteristics with
temperature increase.
\end{abstract}


\maketitle

Silicon nanocrystals (Si-NCs) embedded in a silicon dioxide matrix
have been actively studied due to their potential for
future-generation opto-electronics and
photovoltaics\cite{priolo2014silicon}. One of the advantageous
features of Si-NCs is a significant dependence of optical properties
on the nanocrystal size, which can provide tunability across a wide
optical range. 
At the same time this leads
to demand for Si-NC size control\cite{Zacharias2002,Laube2014,Han2016}
for device applications.

A typical PL spectrum of a Si-NC ensemble exhibits a broad peak with
the width between 100 and 400 meV
even at helium temperatures\cite{Kanzawa1997,TakeokaPRB2000,Meier2007}, 
whereas single Si-NCs at low
temperatures demonstrate narrow a few-meV-wide emission
lines\cite{Sychugov2005PRL}. It may be expected that the PL peak
broadening is related to the NC size
dispersion\cite{Yorikawa1997APL,Trwoga1998,Meier2007} 
through
the size dependence of the optical-transition energies. However,
finding the exact relation between NC size and optical-transition
energy is a difficult task due to complex structural properties of
Si-NCs in silicon dioxide surrounding\cite{Daldosso2003,Luppi2005PRB}.
Additionally, the mechanism of the radiative recombination in Si-NCs
is not yet thoroughly
understood\cite{Delerue1993,Luppi2005PRB,Godefroo2008,BoerYassPod2012}.

In this Letter we develop a general phenomenological theory of
inhomogeneous broadening in zero-dimensional systems which does not
relay on a particular microscopic model of choice, and
we apply it to interpret our experimental results on Si-NC ensembles.

Let us suppose that there is a one-to-one correspondence between the
energy $E$ of a particular optical transition and NC size
parameter $d$, which is represented by a smooth function $E=f\left(d\right)$. If
the NC size dispersion in an ensemble is described by a distribution
function $P(d)$ then the density of optical transitions (DoOT) per
energy interval $\rho \left( E \right)$ is proportional
to\cite{Yorikawa1997APL}
\begin{equation}
\rho \left( E \right)\propto P\left( f^{ - 1}\left( E \right)\right)
\left| {\frac{{df^{ - 1}\left( E \right)}}{{dE}}} \right|,
\label{Eq:rhoGen}
\end{equation}
where $d=f^{ - 1}\left( E \right)$ is the inverse of
$E=f\left(d\right)$.

The function $f\left(d\right)$ is defined by structural and material
properties of nano-objects and is generally unknown. Here, similarly
to our previous work\cite{NikolaevAPL2009}, we presume that in a
certain interval of the NC size variation the microscopic size
dependence can be approximated by a power function\cite{Delerue1993} (PF)
\begin{equation}
E=f(d)=\frac{A}{d^{\gamma}}+\tilde E_{g},\label{Eq:E0depp}
\end{equation}
where $A>0$, $\gamma>0$ and $\tilde E_{g}$ are phenomenological
parameters. In this Letter we consider size dispersion described by
the asymmetric log-normal (LN)
distribution\cite{Yorikawa1997APL,Limpert2001,Lamaestre2006}
\begin{equation}
P_{LN}(d) = \frac{1}{{d\sigma \sqrt {2\pi } }}\exp
\left( { - \frac{1}{{2{\sigma ^2}}}{{\left[ {\ln \left(
{\frac{d}{{\bar d}}} \right)} \right]}^2}}
\right).\label{Eq:LogNorm}
\end{equation}
Here $\bar d$ is the median of the distribution, and $\sigma$
defines the relative standard deviation $\delta d/d_{av}=\sqrt {\exp
\left( {{\sigma ^2}} \right) - 1}$ of the size parameter $d$ from
its average value ${d_{av}} = \bar d\exp \left( {{\sigma ^2}/2} \right)$.
Observation of LN distributions has been reported in many
systems of nano-objects\cite{Lamaestre2006}, including ensembles of
Si-NCs\cite{Yorikawa1997APL,Zacharias2002,Crowe2011,Dogan2013,Laube2014,Han2016,Laube2016}.

Substitution of Eqs. (\ref{Eq:E0depp},\ref{Eq:LogNorm}) into rhs of
Eq. (\ref{Eq:rhoGen}) produces a new normalized asymmetric
distribution ($E>\tilde E_g$)
\begin{equation}
{P_{PF-LN}}( E )
 =\frac{\exp \left( { - \frac{1}{{2\chi^2}}{{\left[ {\ln \left( \frac{E-\tilde E_g}{\bar E-\tilde E_g} \right)} \right]}^2}} \right)}{{{\chi}\left( E-\tilde E_g \right)\sqrt {2\pi } }},
\label{Eq:LogNormPL}
\end{equation}
where $\chi=\gamma\sigma$; $\bar E$ is the median of the
transition-energy distribution directly connected to the median
size $\bar d$ by means of PF Eq. (\ref{Eq:E0depp}): $\bar E=f(\bar d)=A/{\bar d}^\gamma+\tilde E_g$.

Similarly to the size-dispersion parameter $\sigma$ in the LN distribution
Eq. (\ref{Eq:LogNorm}), $\chi$ defines the asymmetry of the peak and
the relative standard deviation of the ``effective quantized
energy'' $\varepsilon=E - \tilde E_g$ from its average value
$\varepsilon_{av}=E_{av}  - \tilde E_g$ through the expression
$\delta\varepsilon/\varepsilon_{av}=\delta E/(E_{av}  -
\tilde E_g)=\sqrt{\exp(\chi^2)-1}$. Here $E_{av}$ is the average
transition energy.

The  full width at half maximum (FWHM) of the peaked distribution
function $P_{PF-LN}$ can be expressed in an analytical form:
 \begin{equation}
 \Delta E = 2\left( {{E_{max }} - {{\tilde E}_g}} \right){\mathop{\rm sh}\nolimits} \left( {\chi \sqrt {2\ln \left( 2 \right)} }
 \right)\label{Eq:FWHM},
 \end{equation}
 where $E_{max}$ is the peak energy (mode) of the $PF-LN$ distribution.
As can be seen from this expression, the peak broadens with the
increase of the parameter $\chi$. Since $\chi$ is the product of
$\sigma$ (which characterizes the relative width of a size
distribution) and the exponent $\gamma$, we naturally arrive
at the conclusion that the width and the asymmetry of the
DoOT peaks depend on the size dispersion
as well as on the character of the transition-energy dependence on
size. We note that $P_{PF-LN}(E)$
Eq. (\ref{Eq:LogNormPL}) approaches the Gaussian distribution when
$\chi\to 0$.

Eq. (\ref{Eq:FWHM}) can be rewritten in the following form
\begin{equation}
\Delta E = 2\exp \left( { - {\chi ^2}} \right){\mathop{\rm
sh}\nolimits} \left( {\chi \sqrt {2\ln \left( 2 \right)} }
\right)\frac{A}{{{{\bar d}^\gamma }}}\\, \label{Eq:FWHMwav}
\end{equation}
to reveal that, at constant level of size dispersion set by
$\sigma$, the FWHM of an inhomogeneously-broadened optical
transition will increase with the decrease of the median NC size
$\bar d$, and the character of FWHM dependence on size will follow the power
law described by the same exponent $\gamma$ which defines the transition
energy dependence.

We argue that the DoOT in the form of Eq. (\ref{Eq:LogNormPL}) can
be used to model the experimental PL spectra of silicon nanocrystals. 
Indeed, if the
redistribution of electrons and holes between NCs is prohibited by
high potential barriers, variations of the pump-radiation absorption
and  internal quantum efficiency\cite{Miura2006PRB} across an
ensemble are slow then the PL intensity at a certain photon energy
will be defined by the number of NCs with the related
optical-transition energy, i.e. PL spectra will replicate the shape
of the DoOT. In this work we neglect the dependence of the internal
quantum efficiency on size justifying it by the results of
experimental investigations which show that, despite the strong
reduction of the radiative lifetime with the NC size decrease
evident from time-resolved
experiments\cite{TakeokaPRB2000,Miura2006PRB}, internal quantum
efficiency varies moderately, especially in the long wavelength
range for the samples with relatively large NCs\cite{Miura2006PRB}.
\begin{figure}
\includegraphics[scale=1.0]{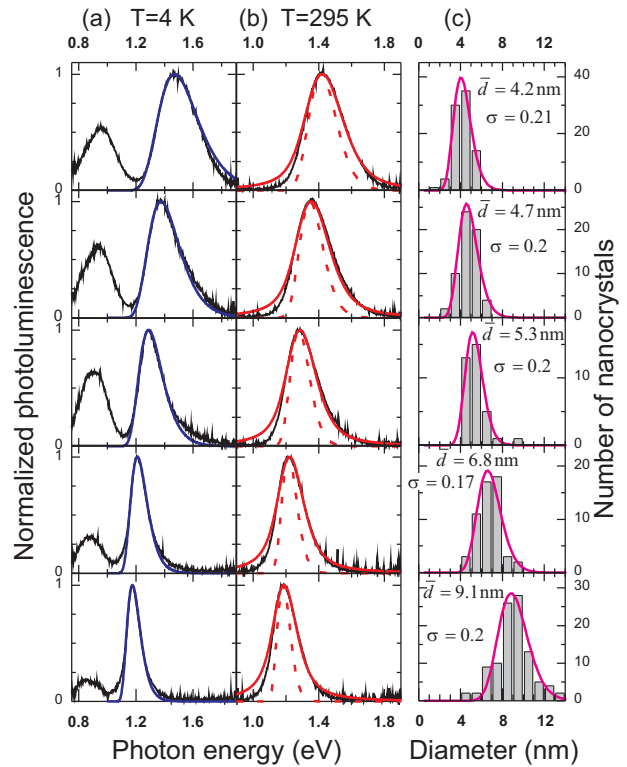}
\caption{(Color online)
Measured and calculated PL spectra at T=4 K (a) and T=295 K (b)
from five samples with different NC size distributions (shown in (c)).
Spectra are normalised to their maximum intensities;
dashed curves in (b) show modelled RT spectra without homogeneous 
broadening.
\label{Fig:SpecSD}}
\end{figure}

In Fig. \ref{Fig:SpecSD} PL spectra measured at helium and room temperatures (RT)
are displayed
alongside diameter
distributions in five Si-NC/$\rm SiO_2$ samples under investigation.
The PL experiments were conducted at a low excitation
power density of 4 $\rm mW/cm^2$, in order to avoid NC ground level saturation and related effects.
Experimental histograms, obtained by high-resolution transmission electron microscopy (TEM),
 were fitted by scaled LN distributions (see Fig. \ref{Fig:SpecSD}(c));
the size-dispersion parameter $\sigma$ was found close to $\sigma_{\rm TEM}\approx 0.2$ in all five samples.
Information on the sample preparation and experimental setup can be
found in Ref. \onlinecite{TakeokaPRB2000}.

The measured 
temperature dependence of PL decay time 
for these samples\cite{TakeokaPRB2000}
was successfully described by the two-level 
Calcott's model\cite{Calcott1993}. The model assumes that PL at low temperatures 
originates from a ground ``semi-dark" exciton level, which
is predominantly spin-triplet with
moderate admixture of spin-singlet character facilitated by 
the spin-orbit interaction. Our measurements at 4 K (Fig.\ref{Fig:SpecSD}(a)) 
show an additional peak associated
with $P_b$ centers at Si/$\rm SiO_2$
interface\cite{Fujii2000JAP}.
At RT PL is dominated by recombination of an excited ``semi-bright"
exciton level which is split from ``semi-dark" level by 
the electron-hole exchange interaction\cite{Calcott1993} and has mainly spin-singlet
character.    
The main PL peaks at both temperatures
demonstrate significant degree of asymmetry alongside with the
blueshift and linewidth broadening with the decrease of NC
size.

Measurements on single Si-NCs show\cite{Sychugov2005PRL} that at
cryogenic temperatures the homogeneous broadening of an emission
line is less than 3 meV and, therefore, at T=4 K it can be safely
neglected compared to peak widths exceeding 100 meV observed in our
experiments. In this case Eq. (\ref{Eq:LogNormPL}) and related
equations can be used directly to model experimental data at T=4 K,
provided the assumptions listed above are valid.

\begin{figure}
\includegraphics{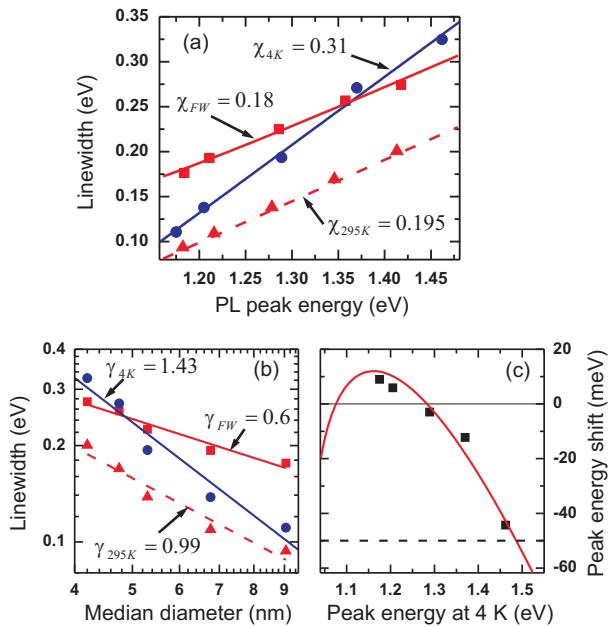}
\caption{(Color online) {\bf (a,b)} PL  FWHM measured at T=4 K
(blue circles), T=295 K (red squares) and extracted inhomogeneous
linewidths  for T=295 K (red triangles) as functions of PL peak energy $E_{max}$ (a) and
NC median diameter $\bar d$ (b) (note log-log scale in (b)).
{\bf (c)} Energy shift of the PL peak $E_{max}$ with the temperature
increase from 4 K to 295 K as function of $E_{max}$ at 4 K.
Lines in (a-c) represent modelling. \label{Fig:FWHM}}
\end{figure}

Figure \ref{Fig:FWHM}(a) shows the FWHM $\Delta E$ of the main PL peaks from
Fig. \ref{Fig:SpecSD}(a,b) as functions of peak energies $E_{max}$. Experimental
points lay close to straight lines. Analysis of Eq. (\ref{Eq:FWHM})
reveals that linear dependence can take place when PF Eq. (\ref{Eq:E0depp}) describes transition energies in all
five samples and NC diameters are distributed log-normally with an
approximately equal value of the size-dispersion parameter $\sigma$. The values of
phenomenological parameters for T=4 K $\tilde E_{g4K}=1.026\pm0.013$
eV and $\chi_{4K}=0.31\pm0.02$ are obtained straightforwardly
 by fitting the linear dependence
Eq. (\ref{Eq:FWHM}) to the experimental data. These values were used
to model the PL spectra with the scaled functions
$P_{PF-LN}(\bar E, \chi_{4K}, \tilde E_{g4K}; E=\hbar\omega )$ , Eq.
(\ref{Eq:LogNormPL}), where median energies $\bar E$ (which are
close to the PL peak energies) are found by fitting.
In Fig. \ref{Fig:SpecSD}(a) a good agreement between calculated and experimental data
can be seen, with the modelled spectra perfectly replicating the
asymmetry and the linewidth variation of the measured PL peaks.

In Fig. \ref{Fig:FWHM}(b) FWHM values are plotted as functions of the NC median
diameter. According to our model (see Eq.
(\ref{Eq:FWHMwav})) this dependence should be close to linear in
log-log scale, and the slope of the line gives the value of the
exponent ($\gamma_{4K}=1.4\pm0.2$ for T=4 K). 
The extracted parameters allow us to estimate the level of
size dispersion from PL spectra alone 
$\sigma=\chi_{4K}/\gamma_{4K}\approx 0.22$
approximately matching TEM results.
Additionally, the dependence of the retrieved 
median effective quantized energies $\bar\varepsilon=\bar E-\tilde E_g$
on median diameter, plotted in log-log scale (not shown), is approximately linear, 
which is in accordance with Eq. (\ref{Eq:E0depp}); 
and the obtained absolute value of the slope $1.4\pm0.2$
equals the previous result for $\gamma_{4K}$.
These findings suggest that the presented theoretical approach
is consistent with the experimental results at T=4 K.

The retrieved value of exponent $\gamma_{4K}$  agrees well with
$\gamma=1.39$ calculated by Delerue\cite{Delerue1993} {\it et
al.}. The effective bandgap
$\tilde E_{g4K}\approx1.025$ eV is smaller than the
bandgap of crystalline Si (1.17 eV),
which is possibly partly due to
an effective hydrostatic pressure\cite{Zhu1989} related to
the NC surface\cite{Crowe2011},
the energy of an emitted optical
phonon\cite{Fujii2005PRB} and the bulk-exсiton binding energy. 
In addition, $\tilde E_{g4K}$ may differ from the value of the material
bandgap because the formula Eq. (\ref{Eq:E0depp}) 
approximates the transition energy dependence only
in a limited diameter range.

Comparing the PL spectra measured at 4 K and 295 K one may find some
unexpected effects. Firstly, the FWHM growth with the increase of
PL peak energy (Fig. \ref{Fig:FWHM}(a)) or decrease of NC size (Fig. \ref{Fig:FWHM}(b)) is
less steep at RT. More surprisingly, the two samples
with the smallest NCs ($\bar d=4.2$ and 4.7 nm) demonstrate PL
peak narrowing with the increase of the temperature (see Fig. \ref{Fig:FWHM}(b)).
The shift of the PL peak energy with the temperature increase is
plotted in Fig. \ref{Fig:FWHM}(c) as function of the peak energy at 4 K. One can
see that the two samples with smallest PL peak energies (and largest
median NC diameters $\bar d=6.8$ and 9.1 nm) exhibit blueshift with
the temperature increase. This is a counter-intuitive behaviour since
the corresponding Si bandgap reduction is approximately 50 meV
(shown by a dashed line in Fig. \ref{Fig:FWHM}(c)) Saturation of NC ground
levels at 4 K cannot serve as an explanation for such blueshift,
since it has the opposite effect\cite{Hartel2012PRB}.

An attempt to directly apply the approach used for T=4 K to the
experimental results at T=295 K leads to the following contradiction.
Fitting the linear dependence Eq. (\ref{Eq:FWHM}) to the
experimental FWHM in Fig. \ref{Fig:FWHM}(a) gives $\chi_{FW}=0.18\pm0.01$ and 
an unexpectedly low value for
the effective bandgap $E_{gFW}=0.75\pm0.03$ eV; application of Eq. (\ref{Eq:FWHMwav}) to
the measured data (red squares in Fig. \ref{Fig:FWHM}(b)) produces the value
of exponent $\gamma_{FW}=0.6\pm0.08$ leading to the estimate of the
size-dispersion parameter $\sigma=\chi_{FW}/\gamma_{FW}\approx 0.3$
which exceeds the experimentally obtained $\sigma_{\rm TEM}\approx0.2$ by
half.

This contradiction occurred because homogeneous broadening of
Si-NC optical transitions was not accounted for by the analysis above.
Experiments show\cite{Sychugov2005PRL,Valenta2002APL} that single
Si-NCs exhibit quite large  PL linewidths (100-150 meV) at RT.
Manifestation of homogeneous broadening of more than 100 meV was also
seen in ensemble investigations\cite{Fujii2005PRB}.

To account for homogeneous broadening we convolute DoOT given by $P_{PF-LN}(E)$
Eq. (\ref{Eq:LogNormPL}) with the Lorentzian
function $L(E)=2\Gamma_{hom}/\left[\pi\left(4E^2+\Gamma_{hom}^2\right) \right]$,
where $\Gamma_{hom}$ is the FWHM of a single-NC transition.
It was found that the
best fit to experimental data can be achieved if $\Gamma_{hom}=130$
meV is used, which is in accordance with reported
measurments\cite{Valenta2002APL}. The parameters
$\chi_{295K}\approx0.195$ and $\tilde E_{g295K}\approx0.98$ eV were
chosen to match the measured RT FWHM in Fig. \ref{Fig:FWHM}(a). The resulting
spectra simulations are shown in Fig. \ref{Fig:SpecSD}(b) demonstrating good
agreement with experiment. Slight overestimation of PL intensity
in the low-photon-energy region might be due to
the deviation of the actual size dependence from PF Eq. (\ref{Eq:E0depp})
for larger NCs.

The value of the exponent $\gamma_{295K}=0.99\pm0.13$ can be derived
from the dependence of the obtained inhomogeneous part of 
the linewidth on NC size (Fig. \ref{Fig:FWHM}(b), red triangles). The size
dependence of the median effective quantized energies
$\bar\varepsilon=\bar E-\tilde E_{g295K}$ (not shown) produces the
same value of exponent. 
The ratio 
$\chi_{295K}/\gamma_{295K}\approx 0.2$
coincides with $\sigma_{\rm TEM}$,
which shows that
our model is consistent.

Now an explanation to the anomalous PL linewidth narrowing and
blueshift with temperature increase can be proposed. With elevation
of the temperature from 4 K to 295 K the dominant luminescence channel
switches from recombination of semi-dark lower exciton states to upper semi-bright
states, and the energy dependence on size of semi-bright states at RT
differs from that of semi-dark states at 4 K. Both dependencies can be approximated by
Eq. (\ref{Eq:E0depp}) but with remarkably different parameters.
Effective bandgap $\tilde E_g$ shrinks from approximately 1.025 eV
to 0.98 eV, which is in line with the Si bandgap reduction by 50
meV. The other two phenomenological parameters change as follows:
$\gamma_{4K}\approx 1.4$ to $\gamma_{295K}\approx 1.0$ and
$A_{4K}\approx3.3$ ${\rm eVnm^{1.4}}$ to $A_{295K}\approx1.8$ eVnm
(values of $A$ are obtained by fitting Eq. (\ref{Eq:FWHMwav}) to
experimental data in Fig. \ref{Fig:FWHM}(b)).

Decrease in the value of the exponent $\gamma$ means that the
variation of the optical-transition energy with the NC size becomes
slower, which leads to the reduction of the inhomogeneous broadening
with temperature increase (compare blue and dashed red lines in
Figs. \ref{Fig:FWHM}(a,b)). The thermal broadening of a single-NC transition
leads to the increase of the total peak width by 70-80 meV, 
which partly compensates the decrease of inhomogeneous broadening. For samples
with small NC sizes the decrease of inhomogeneous broadening is most
pronounced, and the homogeneous broadening is unable to compensate
it, resulting in the total linewidth narrowing with temperature
elevation.

We calculated the difference between PL peak maxima at 4 K and 295 K
using phenomenological parameters listed above (continuous red line
in Fig. \ref{Fig:FWHM}(c)); the size-dispersion parameter was set to
$\sigma=0.2$. One can see that the modelling reproduces the anomalous
blueshift for structures with larger NC diameters. The origins of
this effect are as follows. Firstly, the semi-bright state at 295 K
approaches bandgap with the size increase more slowly than the
semi-dark state at 4 K, which is due to $\gamma$ reduction. At certain
interval of (comparatively large) NC diameters semi-bright transition
energies at T=295 slightly exceed semi-dark transition energies at 4 K
despite the reduced RT effective bandgap $\tilde
E_{g295K}$ (45 meV smaller than $\tilde E_{g4K}$). Secondly,
enhancement of thermal broadening blueshifts the PL peak  by 5-10
meV.

Significant reduction of $\gamma$ with temperature rise indicates that
physical properties of Si-NCs in $\rm SiO_2$ matrix strongly differ
from those of homogeneous crystalline Si. The mechanism which
leads to $\gamma$ alteration may be due to
complex effects which temperature elevation imposes on
stress profiles\cite{Kleovoulou2013PRB}, the transition region between a Si core and
$\rm SiO_2$ matrix\cite{Daldosso2003,Luppi2005PRB} or defect/surface states\cite{Godefroo2008},
and still needs to be clarified.
It should be noted that, according to experimental
observations, for smaller Si-NCs in $\rm SiO_2$ matrix the blueshift
with the size reduction is quenched and the PL peak energy normally
does not exceed 1.8-2.0 eV. From this one may conclude
that the divergent PF Eq. (\ref{Eq:E0depp}) will not
adequately describe NCs with small diameters and, therefore, our
method is not applicable to such samples.

In conclusion, employing log-normal size
distribution\cite{Limpert2001,Lamaestre2006} we developed a concise
phenomenological theory of inhomogeneous broadening in
zero-dimensional systems which contains few fitting parameters.
Obtained analytical expressions allow for relatively straightforward
interpretation of experimental results. The proposed approach
clarifies the measured line shape and linewidth variations of the PL
spectra of silicon nanocrystals. We explain anomalous PL line
narrowing and blueshift with temperature increase by 
competition between opposite variations of 
homogeneous and inhomogeneous broadening and by modification
of the transition-energy size dependence.

\bibliography{SiNCbib}

\end{document}